\documentclass[letterpaper, 10 pt, conference]{ieeeconf}  

\IEEEoverridecommandlockouts                              

\usepackage{graphicx}
\usepackage{amsmath,amssymb,amsfonts}
\usepackage{multirow}
\usepackage{psfrag}
\usepackage{subfigure}
\usepackage{color}
\usepackage{soul}
\usepackage{xcolor}
\usepackage[hyphens]{url}
\usepackage[bookmarks=false]{hyperref}
\usepackage[hyphenbreaks]{breakurl}
\usepackage{lipsum}
\usepackage[ruled,vlined,linesnumbered]{algorithm2e}
\usepackage{mathrsfs}
\usepackage{pbox}
\usepackage{wrapfig}
\usepackage{array}

\DeclareMathOperator*{\argmaxA}{max}

\makeatletter
\newcommand{\thickhline}{%
    \noalign {\ifnum 0=`}\fi \hrule height 1pt
    \futurelet \reserved@a \@xhline
}
\newcolumntype{"}{@{\hskip\tabcolsep\vrule width 1pt\hskip\tabcolsep}}
\makeatother

\DontPrintSemicolon

\makeatletter
\def\hlinewd#1{%
  \noalign{\ifnum0=`}\fi\hrule \@height #1 \futurelet
   \reserved@a\@xhline}
\makeatother

\graphicspath{{Figures/}}

\title{\LARGE \bf
D-ACC: Dynamic Adaptive Cruise Control for Highways with Ramps Based on Deep Q-Learning
}

\author{Lokesh Das and Myounggyu Won\\
Department of Computer Science, University of Memphis, Memphis, TN, United States\\
\{ldas, mwon\}@memphis.edu}

\begin{document}

\maketitle
\thispagestyle{empty}
\pagestyle{empty}


\begin{abstract}
An Adaptive Cruise Control (ACC) system allows vehicles to maintain a desired headway distance to a preceding vehicle automatically. It is increasingly adopted by commercial vehicles. Recent research demonstrates that the effective use of ACC can improve the traffic flow through the adaptation of the headway distance in response to the current traffic conditions. In this paper, we demonstrate that a state-of-the-art intelligent ACC system performs poorly on highways with ramps due to the limitation of the model-based approaches that do not take into account appropriately the traffic dynamics on ramps in determining the optimal headway distance. We then propose a dynamic adaptive cruise control system (D-ACC) based on deep reinforcement learning that adapts the headway distance effectively according to dynamically changing traffic conditions for both the main road and ramp to optimize the traffic flow. Extensive simulations are performed with a combination of a traffic simulator (SUMO) and vehicle-to-everything communication (V2X) network simulator (Veins) under numerous traffic scenarios. We demonstrate that D-ACC improves the traffic flow by up to 70\% compared with a state-of-the-art intelligent ACC system in a highway segment with a ramp.
\end{abstract}

\section{Introduction}
\label{sec:introduction}

An adaptive cruise control (ACC) is a control system that allows vehicles to keep a desired headway distance to the front vehicle by adjusting acceleration and deceleration using radar, lasers, or cameras~\cite{vahidi2003research}. More and more vehicles are equipped with ACC, increasing the market penetration rate every year~\cite{manolis2020real}. It is becoming standard equipment in many recently available commercial vehicles~\cite{makridis2020empirical}. While ACC is primarily designed for improving driving comfort, numerous research has been performed to demonstrate its benefits on improving traffic flow~\cite{davis2004effect}. A general consensus of these works based on microscopic simulation~\cite{ntousakis2015microscopic, bayar2016impact} or macroscopic modelling~\cite{delis2016simulation,nikolos2015macroscopic} is that ACC has strong potential to improve traffic flow, and poor configurations of ACC settings may result in degraded traffic flow~\cite{delis2016simulation}. However, most of those works focus only on demonstrating the effect of ACC on traffic flow and do not provide an adaptive solution that configures ACC settings to maximize traffic flow in response to dynamically changing traffic conditions.

There are some efforts to optimize ACC settings to maximize traffic flow~\cite{mba2016evaluation,liu2017fine}. However, these works are based on off-line optimization which does not cope well with dynamically changing traffic conditions. There are several methods that address this limitation by adjusting the ACC setting adaptively in real time according to dynamic traffic conditions~\cite{schakel2014improving,bekiaris2019feedback,goni2019using}. Specifically, Shakel and Arem~\cite{schakel2014improving} develop an in-car advisory system that changes the headway distance based on a traffic state prediction model (\emph{i.e.,} the extended generalized Treiber–Helbing filter (EGTF))~\cite{van2010robust}. This model determines between free flow and congestion based on the speed data collected from a detector that are aggregated over 1 min. Goni-Ros \emph{et al.} develop an adaptive ACC that adjusts the ACC setting based on the speed of preceding vehicles~\cite{goni2019using}. Bekiaris-Liberis and Delis propose a similar approach that changes the ACC setting in real time based on the average speed and vehicle density~\cite{bekiaris2019feedback}. Spiliopoulou \emph{et al.} design a threshold-based adaptive ACC that adjusts the ACC setting of a connected vehicle based on a threshold of traffic flow measured using the speed of surrounding vehicles~\cite{spiliopoulou2018adaptive}. In line with this research, the most recent study at the point of writing this paper is~\cite{manolis2020real}. Essentially, this paper extends~\cite{spiliopoulou2018adaptive} by adding acceleration bound changes considering that latest vehicles allow drivers to choose the acceleration strength inspired by Yuan \emph{et al.}~\cite{yuan2016capacity}. While these model-based approaches enable real-time adaptation of ACC settings according to the current traffic conditions, such static models do not represent effectively the traffic dynamics especially for general highways with ramps (including on/off ramps).

\begin{figure*}[h]
	\centering
	\includegraphics[width=\textwidth]{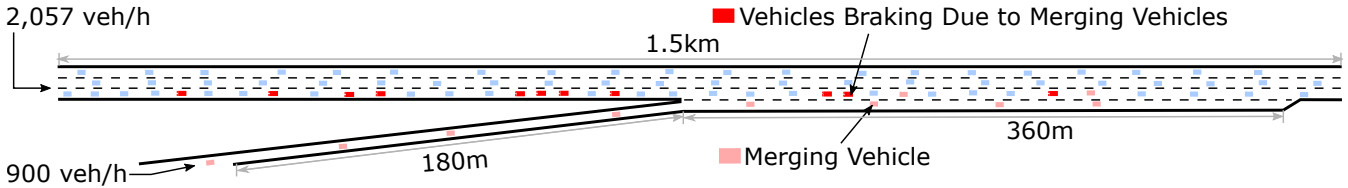}
	\caption{A snapshot of simulation demonstrating vehicles equipped with a state-of-the-art real-time ACC~\cite{manolis2020real} interfering with merging traffic. \label{fig:simulation_snapshot}}
\end{figure*}

\begin{figure*}[h]
	\centering
	\includegraphics[width=\textwidth]{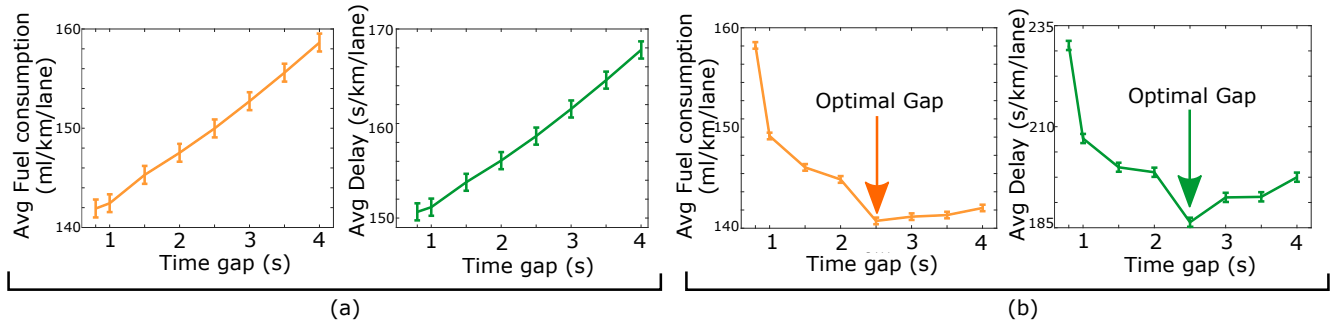}
	\caption{Average fuel consumption and average delay (a) without merging traffic, (b) with merging traffic. \label{fig:simulation_result}}
\end{figure*}

In this paper, we propose a novel AI (Artificial Intelligence)-based approach called the Dynamic Adaptive Cruise Control (D-ACC) System for real-time adaptation of ACC settings in response to dynamically changing traffic conditions. To enable vehicles to adapt the ACC setting more effectively in a fine-grained manner according to the current traffic conditions, we adopt deep reinforcement learning (RL) which is known to enable effective decision making in such a complex environment especially for autonomous vehicles~\cite{chen2017end, triest2020learning}. Specifically, we formulate the problem of determining the optimal ACC setting based on the traffic condition information received via vehicle-to-everything communication (V2X) as a Markov Decision Process (MDP) framework~\cite{sutton2018reinforcement}. We then solve the problem by designing a deep Q-network~\cite{mnih2013playing} that addresses the challenge of dynamic adaptation of the ACC setting in complex environments represented by a large and continuous state space. In particular, considering that merging/exiting traffic is one of the major causes of traffic perturbations on highways~\cite{gupta2006phase, van2017impacts}, we demonstrate the high adaptability of D-ACC under general highway scenarios with on/off ramps. Extensive simulations are conducted using a combination of a microscopic road traffic simulator SUMO~\cite{fernandes2010platooning} and a V2X network simulator Veins~\cite{sommer2010bidirectionally} to train the deep Q-network and evaluate the performance of D-ACC. We demonstrate that D-ACC improves traffic flow represented by the average vehicle speed by up to 70\% compared to a state-of-the-art real-time ACC system~\cite{manolis2020real}. The contributions of this paper are summarized as follows.

\begin{itemize}
  \item We propose a novel AI-based ACC system that adaptively configures the ACC settings in response to dynamically changing traffic conditions, overcoming the limitations of the static model-based approaches.
  \item We formulate the problem of dynamic adaptation of ACC settings as a MDP framework in reference to the transportation literature. A deep-Q network is designed to allow vehicles to make informed decisions on adapting the ACC setting more effectively under complex and dynamically changing traffic conditions.
  \item Extensive simulations are conducted under general highway scenarios with ramps. We demonstrate that D-ACC outperforms a state-of-the-art intelligent ACC system in various traffic conditions with varying the penetration rates, traffic density of both the ramp and main lane, and length of the acceleration lane, and lane-changing behavior.
\end{itemize}

This paper is organized as follows. In Section~\ref{sec:motivation}, we conduct a motivational study to demonstrate that a state-of-the-art real-time ACC system performs poorly due to not appropriately accounting for the dynamics of traffic on a ramp. We then present the details of the proposed D-ACC including the design of a MDP framework and our deep Q-network for optimizing traffic flow in Section~\ref{sec:proposed_approach}. The simulation settings and results are presented in Section~\ref{sec:results}. Finally, we conclude in Section~\ref{sec:conclusion}.

\section{Motivation}
\label{sec:motivation}


In this section, we demonstrate that a state-of-the-art ACC system does not perform well in determining the optimal ACC setting for highways with ramps. This motivational study is conducted using SUMO~\cite{fernandes2010platooning} and Veins~\cite{sommer2010bidirectionally}. The Veins, a simulation framework for vehicular network based on OMNeT++~\cite{varga2008overview}, is integrated with SUMO to simulate V2X for vehicles to collect traffic information. Specifically, a highway segment with an on-ramp is considered where vehicles adjust their gaps to the preceding vehicle using a state-of-the-art ACC system~\cite{manolis2020real}. All parameters used for this simulation study are summarized in Table II in Section~\ref{sec:sim_setup}. Fig.~\ref{fig:simulation_snapshot} illustrates a snapshot of simulation where merging vehicles are causing traffic perturbations because of the small headway gap suggested by the state-of-the-art ACC system~\cite{manolis2020real} without considering the traffic on the ramp.

To understand why existing intelligent ACC systems do not perform well for highways with ramps, we vary the headway distance and measure the average fuel consumption and average vehicle delay (metrics selected according to~\cite{manolis2020real}) in both scenarios with and without merging traffic. Fig.~\ref{fig:simulation_result}(a) shows the results. When there is no merging traffic, a small headway distance improves fuel efficiency and delay. This is because a smaller headway gap allows more vehicles to pass the highway segment given the same amount of time. These results agree with the findings of state-of-the-art intelligent ACC systems~\cite{spiliopoulou2018adaptive,manolis2020real}.

However, these model-based ACC systems perform poorly when there is merging traffic as shown in Fig.~\ref{fig:simulation_result}(b). It is observed that the fuel efficiency and delay actually degrade as the headway gap decreases. The reason is because a small headway gap makes lane changing for merging vehicles very difficult, and a lane change can cause the following vehicles to apply strong braking, leading to higher fuel consumption and delay. A very interesting observation is that when the headway gap is too large, the fuel efficiency and delay start to degrade, \emph{i.e.,} there is a ``sweet spot'' for the headway gap that maximizes traffic flow, as indicated by the two arrows in Fig.~\ref{fig:simulation_result}(b). We also observe that such a sweet spot changes when the traffic conditions change over time. This motivational study suggests that a new ACC system should be developed that adjusts the ACC setting adaptively by taking into account the dynamically changing traffic conditions of both the main road and ramp.

\section{Dynamic Adaptive Cruise Control (D-ACC)}
\label{sec:proposed_approach}


\subsection{Overview}
\label{sec:overview}

D-ACC dynamically adjusts the headway distance to maximize traffic flow of a highway with a ramp. It can be used by any vehicle equipped with a standard ACC system and V2X capability. Note that although the V2X technology has yet to be widely adopted, D-ACC can be implemented with a smartphone that can communicate with a roadside unit (or a remote traffic server via a cellular network) to collect real-time traffic information needed to determine the headway distance. The smartphone is connected to the vehicle's default ACC system via the OBD port (\emph{e.g.,} Comma.AI~\cite{santana2016learning}) to control the headway distance. The real-world implementation and experimentation of D-ACC is our future work. In this paper, we focus on the system design and verification of performance via simulation.

As a vehicle equipped with D-ACC approaches a highway segment, it starts to communicate with a RSU such as the one that is already available in latest intelligent transportation systems (ITS)~\cite{duret2020hierarchical} via V2X. The RSU is used to collect real-time traffic information and broadcast it to approaching vehicles. In this work, we utilize possibly all relevant information to make a decision for controlling the headway distance including (1) traffic density of the main lanes, (2) average vehicle speed of the main lanes, (3) traffic density of the ramp, (4) average vehicle speed of the ramp, and (5) the length of the ramp. A vehicle receives such information wirelessly and runs D-ACC to determine the optimal headway distance to maximize traffic flow.

\subsection{Markov Decision Process (MDP) Framework}
\label{sec:mdp_framework}

A Markov decision process (MDP)~\cite{sutton2018reinforcement} represents the basis for modeling the decision making behaviors for D-ACC. We formulate the dynamic decision making process of D-ACC as MDP in which D-ACC-equipped vehicles take actions of adjusting the headway distance in response to dynamically changing traffic conditions.

We represent MDP as a 4-tuple $M=<S,A,P_{sa},R>$, where $S$ is a set of states; $A$ is a set of actions; $P_{sa}$ is the probability for the next state given action $a \in A$ and the current state $s \in S$; and $R$ is the reward function, \emph{i.e.,} $r(s'|s,a)$ is the reward for transitioning from a state $s$ to a new state $s'$ due to an action $a$. In particular, $\pi_t(a|s)$ is a policy of time $t$ that represents the probability of making an action $a$ given state $s$. The objective is to find the optimal policy $\pi^*$ that makes the cumulative sum of the expected reward:

\begin{equation}
Q_{\pi}(s,a) = \mathbb{E}_{\pi}[\sum_{k=0}^{\infty}\gamma^kr_{t+k}|s,a]
\end{equation}

\noindent is maximized over the long run, \emph{i.e.,} $Q_{\pi}^*(s,a) = \argmaxA\limits_{\pi} Q_{\pi}(s,a)$, where $\gamma \in (0,1]$ is a discount factor.

\subsubsection{State Space}
\label{sec:state_space}

The state space $S$ of MDP is designed to include the following traffic parameters being motivated by~\cite{daamen2010empirical}: (a) traffic density of the main road, (b) traffic density of the ramp, (c) average vehicle speed of the main road, (d) average vehicle speed of the ramp, and (e) length of the ramp. Specifically, Daamen \emph{et al.} note that the traffic density of the main road as well as the ramp influence the vehicle merging behaviors~\cite{daamen2010empirical}. For example, if the traffic density of the ramp is high, then a larger headway distance would be needed to allow vehicles to join quickly so more vehicles can merge into the main lane. It is also noted in~\cite{daamen2010empirical} that the vehicle speed, in particular, the speed difference between the vehicles on the ramp and the vehicles on the main road influences the efficiency of merging behavior~\cite{daamen2010empirical}. For example, if the speed of the vehicles on the ramp is relatively higher, a large gap should be created to ensure safer lane change. Finally, the length of a ramp is incorporated into our state space because a longer ramp would allow vehicles to take more time to complete merging/exiting maneuvers.


\subsubsection{Action Space}

The action space $A$ of MDP is specifically designed for vehicles in the main lane to adjust the headway distance dynamically to allow vehicles to merge/exit more efficiently and consequently improve traffic flow. More precisely, an action performed by a vehicle is defined as setting the headway distance to a certain value selected from a range of possible headway distances. It is worth to mention that D-ACC allows vehicles to perform an action before they reach a highway segment with a ramp. This is motivated by recent research~\cite{goni2019using} demonstrating that creating a certain inter-vehicle gap before it reaches a congested (or potentially congested) area not only allows vehicles to merge/exit more smoothly but also make the following vehicles decelerate appropriately, reducing the inflow into the congested area and preventing from worsening the traffic congestion.


\subsubsection{Reward Function}

The design of the reward function of MDP is focused on maximizing traffic flow. More specifically, the reward function is designed to decrease the average vehicle delay required to pass a highway segment. Thus, in computing the reward value $R$, the traffic congestion speed of the highway segment $v_{congestion}$ and the length of the highway semgent $l_{segment}$ received from a RSU via V2X are used in computing the reward value as follows.

\begin{equation}
R =
\begin{cases}
      +1 & \mbox{if average delay} \leq \frac{l_{segment}}{v_{congestion}} \\
      -1 & \mbox{if average delay} > \frac{l_{segment}}{v_{congestion}}
   \end{cases}
\end{equation}

\subsection{Deep Q-Network}
\label{sec:deep_q_learning}

The optimal policy $\pi^*$ can be found using the Q-function $Q_{\pi}^*(s,a) = \argmaxA\limits_{\pi} Q_{\pi}(s,a)$. Specifically, the Q-function follows the Bellman optimality equation:

\begin{equation}
Q^*(s,a) = \mathbb{E}[r + \gamma\argmaxA\limits_{a' \in A}Q^*(s',a') | s,a],
\end{equation}

\noindent which allows us to determine the optimal policy by easily representing the Q-function as a tabular form when the state space is discrete and finite~\cite{sutton2018reinforcement}. However, if the state space is large and continuous, it becomes easily untractable~\cite{mnih2013playing}. As such, to account for the large and continuous state space for D-ACC to incorporate wide ranges of vehicle density, speed, and length of a ramp, \emph{etc.}, we adopt the deep Q-network (DQN) algorithm to obtain the optimal policy~\cite{mnih2013playing}. The idea is to approximate the Q-function using a neural network~\cite{al2019deeppool} where the input to the neural network is the states, and the output is the Q-value for each action. Specifically, the approximate value function for deep Q-network is now denoted by $Q(s,a;\theta_i)$ where $\theta_i$ are the weights of the Q-network at $i$-th iteration. The network is trained to update the weights $\theta_i$ with the loss function: $L_i(\theta_i) = \mathbb{E}_M[(r + \gamma\max\limits_{a'}Q(s',a';\theta_i^{-})-Q(s,a;\theta_i))^2]$, where $\theta_i^{-}$ is the parameter of the network at $i$-th iteration, which is updated based on the weights of the Q-network $\theta_i$.

\begin{table}[h]
\centering
\begin{tabular}{ |p{3.8cm}|p{4cm}|  }
 \hline
 \thickhline
 \multicolumn{2}{|c|}{\textbf{Q-Network Parameters}}\\
 \hline
 \thickhline
  Neural network architecture & 4 hidden layers of size 8, 12, 20, and 16 respectively \\
\hline
 Activation functions & Rectified linear units (ReLU) \\
\hline
 Replay buffer size & 20k samples \\
\hline
 ($\gamma, \epsilon_0, \epsilon_{min}$, $\lambda_{decay})$ & (0.95, 1.0, 0.001, 0.9995) \\
\hline
Batch size & 32  \\
\hline
Loss function & Mean square error (MSE) \\
\hline
Optimization method & Stochastic Gradient Descent (SGD) with learning rate 0.001  \\
\hline
Target network update frequency & 1k episodes  \\
\hline
\end{tabular}
\caption{Parameters used for our deep Q-network.}
\label{tab:param_network}
\end{table}


\begin{wrapfigure}{r}{0.6\columnwidth}
\vspace{-10pt}
  \begin{center}
    \includegraphics[width=\linewidth]{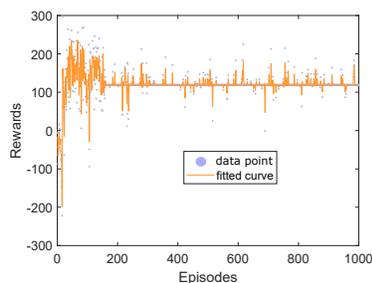}
    \caption{Rewards of the reward function. \label{fig:reward_func}}
  \end{center}
\end{wrapfigure}

For our deep Q-network, we design a neural network consisting of 4 hidden layers of size 8, 12, 20, and 16, respectively with activation functions of rectified linear units (ReLU). Other specific parameters used for designing our deep Q-network through trial and errors are summarized in Table~\ref{tab:param_network}. In particular, to balance between exploration and exploitation, we adopt an $\epsilon$-greedy policy~\cite{wunder2010classes} in training our deep Q-network. Specifically, with the probability of $\epsilon$, an action is randomly selected from the action space; and with the probability of $1 -\epsilon$, the optimal action is selected based on the greedy method. We allow the value of $\epsilon$ to decrease gradually as the algorithm iterates, \emph{i.e.,} $\epsilon = \argmaxA(\epsilon_0 \cdot \lambda_{decay}, \epsilon_{min})$. The parameters for the $\epsilon$-greedy policy are also summarized in Table~\ref{tab:param_network}. Fig.~\ref{fig:reward_func} displays the results of our reward function with the number of episodes up to 1K, demonstrating the fast convergence of the reward function.

\section{Simulation Results}
\label{sec:results}

\subsection{Simulation Setup}
\label{sec:sim_setup}

\begin{table}[h]
\centering
\begin{tabular}{ |p{3cm}|p{4.5cm}| }
 \hline
 \thickhline
 \multicolumn{2}{|c|}{\textbf{Vehicle Parameters}} \\
 \hline
 \thickhline
  Action interval & 1 s \\
\hline
 Vehicle length & 4 m \\
\hline
 Min headway & 2.5 m \\
\hline
 Lane change model & LC2013~\cite{erdmann2015sumo} \\
\hline
 Max acceleration & 2.6 m/s$^2$  \\
\hline
Max deceleration & 4.5 m/s$^2$  \\
\hline
\thickhline
\multicolumn{2}{|c|}{\textbf{Traffic Parameters}} \\
\hline
 \thickhline
 Traffic volume (main lane) & 900$\sim$3700 veh/h (default: 2057 veh/h) \\
 \hline
 Traffic volume (merging lane) & 200$\sim$900 veh/h (default: 900 veh/h)\\
 \hline
 Highway length & 1.5 km \\
 \hline
 Merging lane length & 360 m \\
 \hline
 Acceleration lane length & 50$\sim$180 m (default: 180m) \\
 \hline
\end{tabular}
\caption{Parameters for the simulation.}
\label{tab:param}
\end{table}

A simulation study is performed to evaluate the performance of D-ACC using a traffic simulator (SUMO) integrated with a network simulator (Veins). Simulations are conducted using a PC equipped with a 1.4GHz quad-core Intel Core i5 CPU and 8GB of RAM running on MacOS. Deep reinforcement learning for D-ACC is implemented in Python using Keras and Tensorflow~\cite{abadi2016tensorflow} which is interfaced with SUMO via Traffic Control Interface (TraCI)~\cite{wegener2008traci}.

\begin{figure*}[h]
	\centering
	\includegraphics[width=\textwidth]{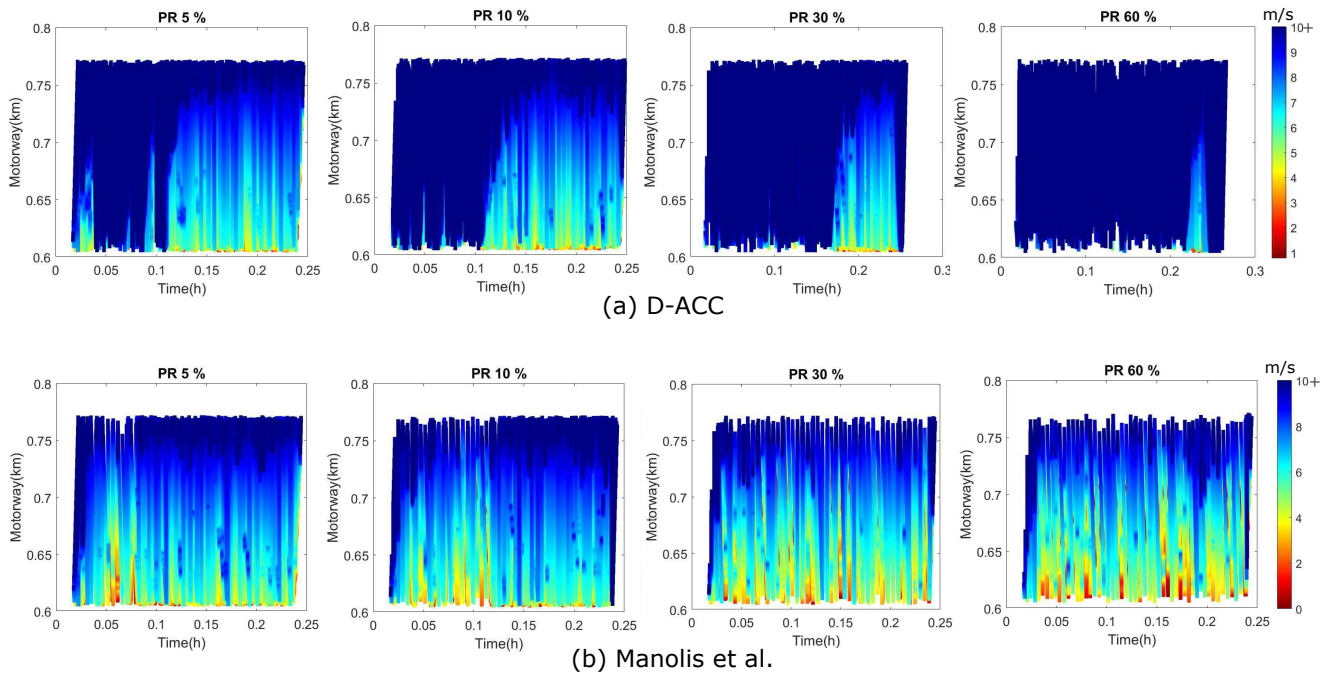}
	\caption{Effect of penetration rate (PR) for (a) D-ACC and (b) Manolis \emph{et al.}~\cite{manolis2020real}. The color represents the vehicle speed in m/s. \label{fig:impact_of_pr}}
\end{figure*}

\begin{figure*}[h]
	\centering
	\includegraphics[width=\textwidth]{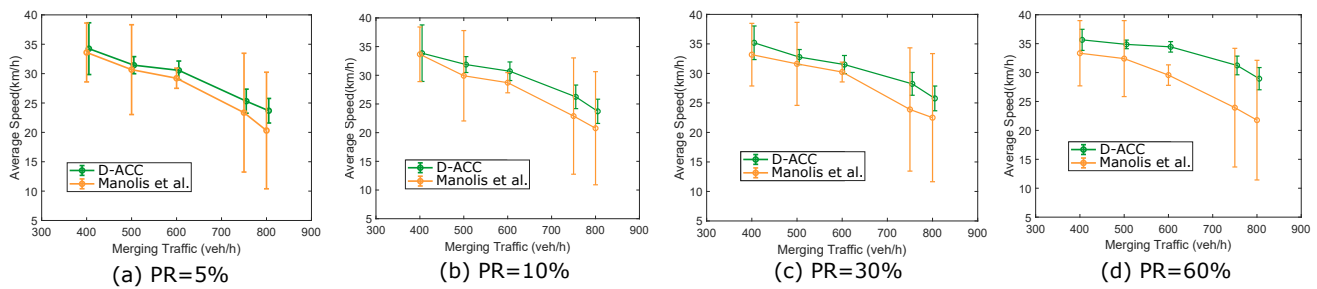}
	\caption{Effect of merging traffic with varying penetration rates. \label{fig:impact_of_merging_traffic}}
\end{figure*}

In this simulation, we consider a highway segment with a ramp. The length of the highway segment, ramp, and acceleration lane are 1.5km, 360m, and (40$\sim$180)m, respectively. Vehicles on the main road are generated at a rate of 900 to 3700 veh/h, and that for the ramp is varied between 200 and 900 veh/h to evaluate the effect of the on-ramp traffic density. Each vehicle with a length of 4m performs a lane change according to the lane-changing model~\cite{erdmann2015sumo}. Each vehicle updates the headway distance every second in response to dynamically changing traffic conditions. We ensure that the headway distance is not decreased below 2.5m for ensuring safety. In increasing or decreasing the headway distance, acceleration of up to 2.6m/s$^2$ and deceleration of up to 4.5 m/s$^2$ are used. All traffic parameters and their default values used for this simulation study are summarized in Table~\ref{tab:param}.

The main metric measured in this simulation study is the average vehicle speed of all vehicles in the given highway segment on a single run that effectively represents the traffic flow. We vary \emph{all possible parameters for the scenario} including the penetration rate (\emph{i.e.} a portion of vehicles equipped with D-ACC or a state-of-the-art real-time ACC out of all vehicles), traffic density of the main lane, traffic density of the ramp, length of the acceleration lane, and lane-changing behavior.

\subsection{Penetration Rate}
\label{sec:penetration_rates}

We measure the average vehicle speed for both D-ACC and Manolis \emph{et al.}~\cite{manolis2020real} by varying the penetration rate. The results are depicted in Fig.~\ref{fig:impact_of_pr} which demonstrate that higher penetrate rates lead to better traffic flow for D-ACC. Interestingly, on the other hand, the opposite results are observed for Manolis \emph{et al.} The reason is that according to Manolis \emph{et al.}, higher penetration rates mean more vehicles trying to maintain smaller headway distance since Manolis \emph{et al.} improve traffic flow by allowing more vehicles to pass the highway segment by using a smaller inter-vehicle gap, which makes merging vehicles difficult to change lanes causing traffic perturbations. Another interesting observation is that D-ACC improves traffic flow by 21\% even with a very small penetration rate of 5\% compared with Manois \emph{et al.}. More significant improvement in traffic flow is observed with higher penetration rates, \emph{i.e.,} over 30\%. Specifically, D-ACC achieves 70\% higher average speed compared with Manolis \emph{et al.} when the penetration rate is 60\%. In other words, the average speed for D-ACC is increased by 11km/h compared with Manolis \emph{et al.} when the penetration rate is increased from 5\% to 60\%. The reason is that with higher penetration rate, more vehicles adjust gaps dynamically to allow merging vehicles to change lanes more smoothly.

\subsection{Traffic Density of Ramp}
\label{sec:merging_traffic_flow}

We evaluate the effect of the traffic density of the ramp on the performance of D-ACC. We measure the average vehicle speed by varying the number of vehicles injected into the merging lane per hour and the penetration rates. The results are depicted in Fig.~\ref{fig:impact_of_merging_traffic} which demonstrate that as the traffic density of the ramp increases, the average speed for both D-ACC and Manolis \emph{et al.} decreases regardless of the penetration rate. An interesting observation is that D-ACC is more resilient to the increasing traffic density of the ramp than Manolis \emph{et al.} as demonstrated by the increasing gap between the average speed of D-ACC and that of Manolis \emph{et al.} when the traffic density increases for all penetration rates. Most notably, the performance gain increases by 210\% as the merging traffic increases from 400 veh/h to 800 veh/h, when the penetration rate is 60\%. It is also observed that the advantages of using D-ACC increase as both the merging traffic and penetration rate increase. Specifically, the gap between the average speed of D-ACC and that of Manolis \emph{et al.} for PR=60\% with merging traffic of 800 veh/h is about 140\% higher compared with the gap for PR=5\% with 800 veh/h. Overall, D-ACC outperforms the state-of-the-art real-time ACC system regardless of the traffic density of the ramp for all penetration rates.

\subsection{Effect of Length of Acceleration Lane}
\label{sec:effect_of_lane}

\begin{figure}[!htbp]
\begin{minipage}[b]{0.49\columnwidth}
\centering
\includegraphics[width=\linewidth]{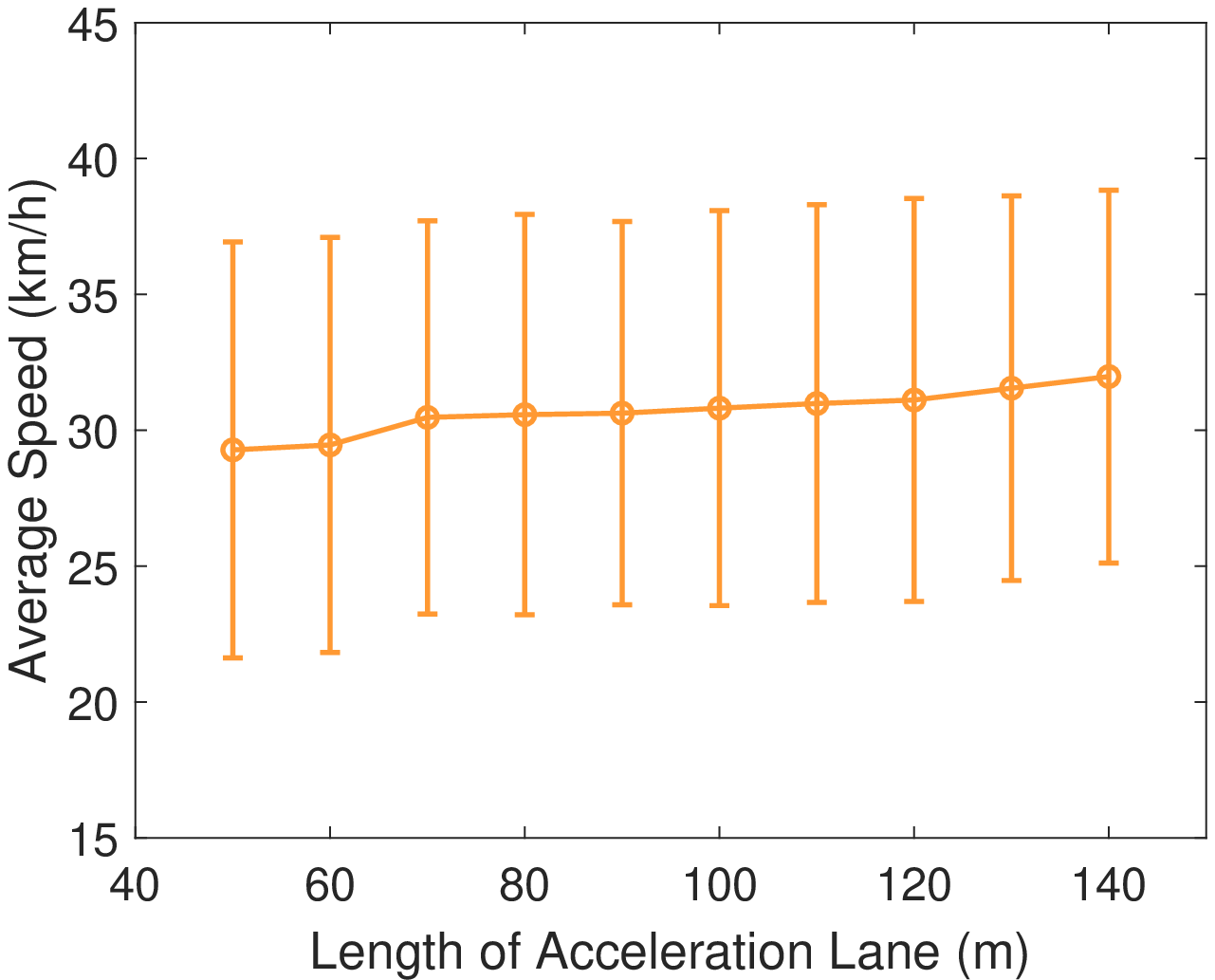}
    \caption{Effect of the length of the ramp. \label{fig:impact_of_lane}}
\end{minipage}
\hspace{0.1mm}
\begin{minipage}[b]{0.49\columnwidth}
\centering
\includegraphics[width=\linewidth]{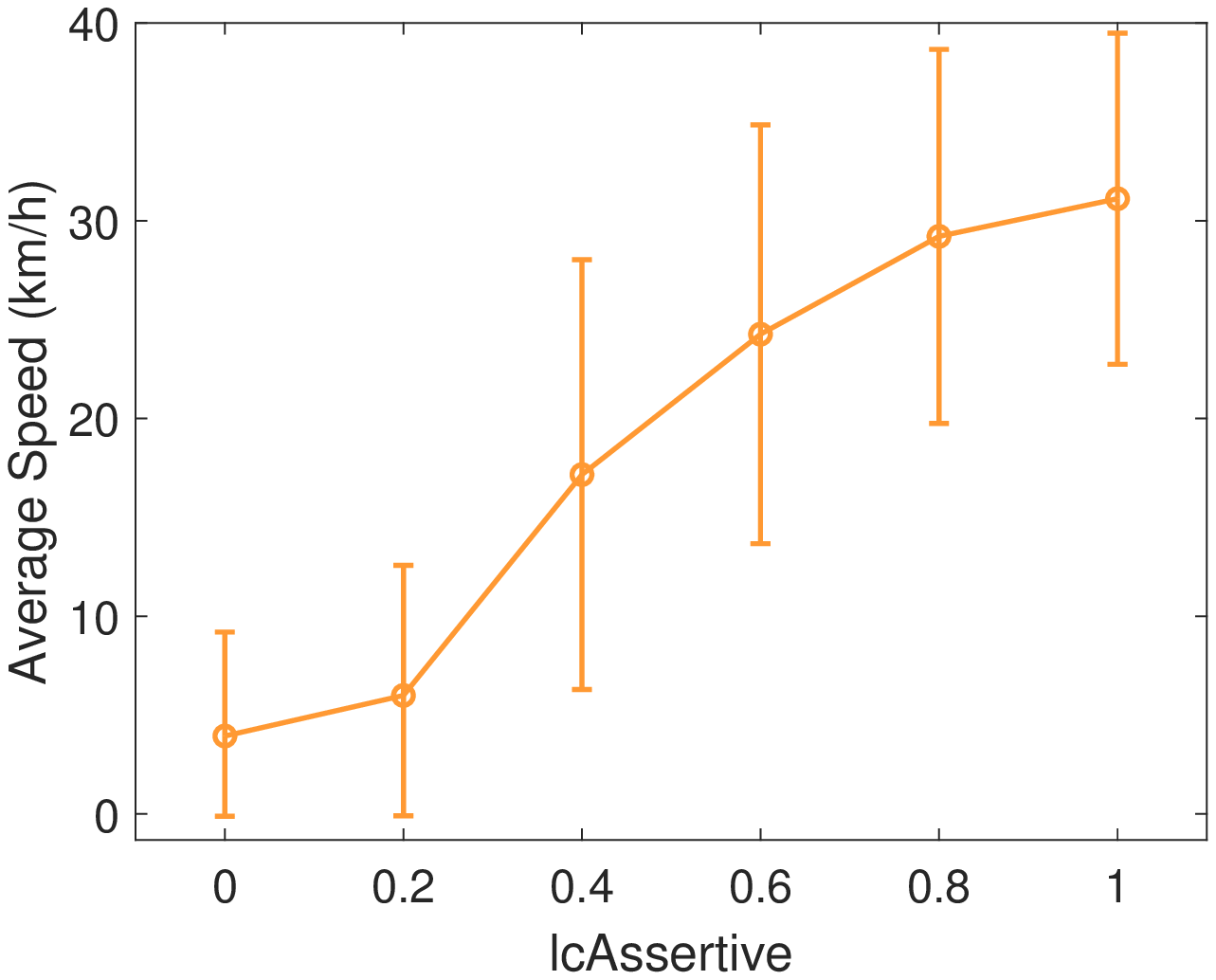}
    \caption{Effect of the lane-changing behavior. \label{fig:impact_of_merging_behavior}}
\end{minipage}
\end{figure}


We evaluate the performance of D-ACC by varying the length of the acceleration lane. We measure the average vehicle speed on the main lane. The results are depicted in Fig.~\ref{fig:impact_of_lane}. We observe that the average speed increases generally as the length of the acceleration lane increases. The reason is because vehicles on the ramp change lanes more easily with a longer acceleration lane, thereby improving overall traffic flow. Specifically, as we increase the length of the acceleration lane from 50m to 140m, the average speed is increased by 9.2\%.

\subsection{Lane Change Behavior}
\label{sec:effect_of_lane_change}


In this section, we analyze the effect of the lane-changing behavior of merging vehicles on the performance of D-ACC. For this simulation, we adopt the default lane-changing model for SUMO~\cite{erdmann2015sumo}. In this lane-changing model, a parameter ``lcAssertive'' is defined to represent the willingness of a driver to accept lower front and rear gaps on the target lane
when executing a lane change~\cite{berrazouane2019analysis}. This parameter is selected between 0 and 1 (with the default value of 1 for SUMO), where a greater number represents more aggressive lane-changing behavior.

We measure the average speed by varying the value of the parameter. The results are depicted in Fig.~\ref{fig:impact_of_merging_behavior}. As shown, D-ACC performs better with larger ``lcAssertive'', \emph{i.e.,} more aggressive lane-changing behavior. The reason is that a larger value of ``lcAssertive'' allows vehicles to change lanes quickly whenever there is a space available to join, thereby giving more time to the following vehicles in the merging lane to make a lane-change decision. On the other hand, a small value of ``lcAssertive'' occasionally make merging vehicles being unable to quickly change the lanes, leading to traffic perturbations, especially when those merging vehicles fail to merge as they reach almost the end of the ramp.

\subsection{Traffic Density of Main Lane}
\label{subsec:main_lane_density}


\begin{figure}[!htbp]
\begin{minipage}[b]{0.49\columnwidth}
\centering
\includegraphics[width=\textwidth]{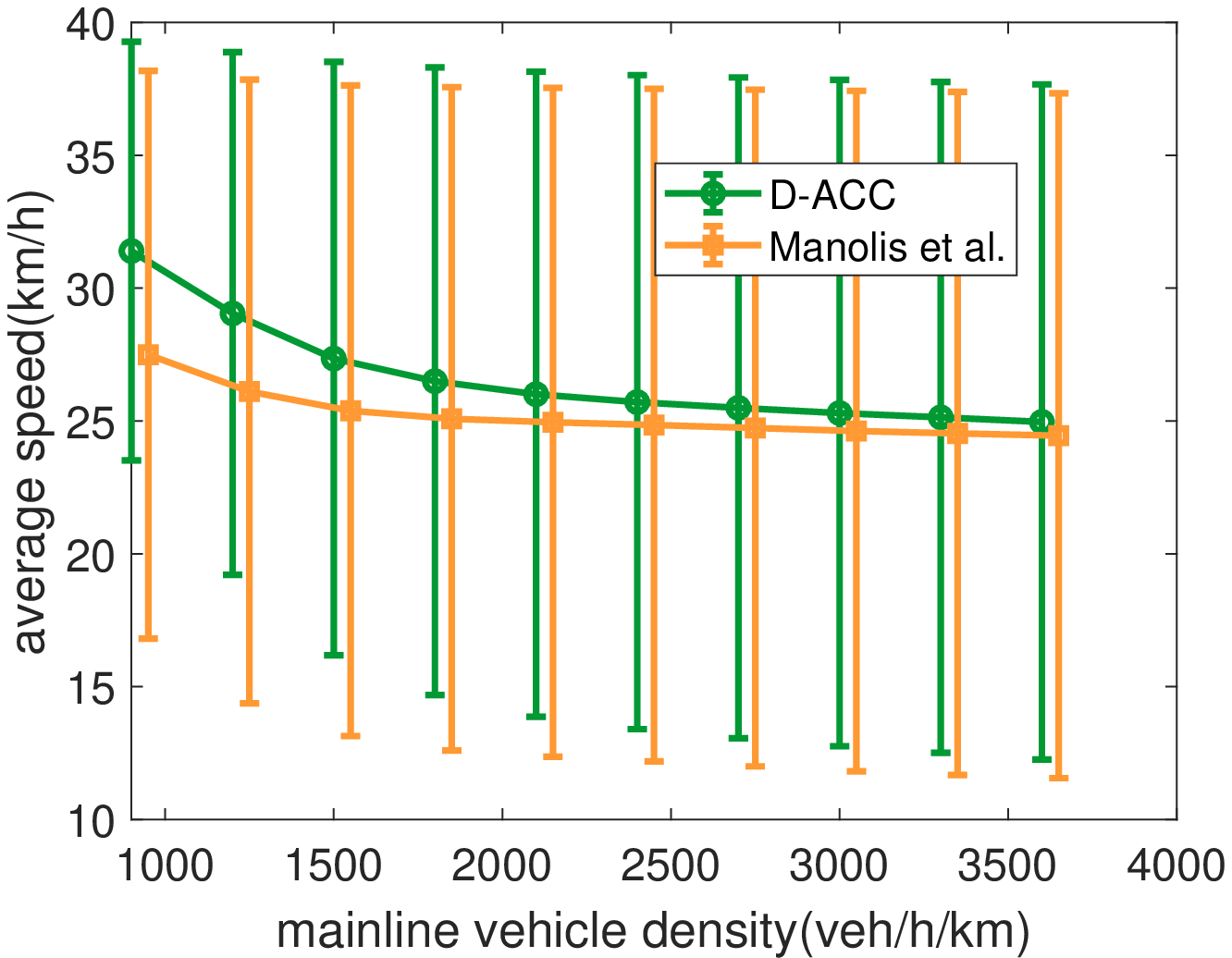}
    \caption{Effect of the traffic density of the main lane. \label{fig:main_lane_density}}
\end{minipage}
\hspace{0.1mm}
\begin{minipage}[b]{0.49\columnwidth}
\centering
\includegraphics[width=\textwidth]{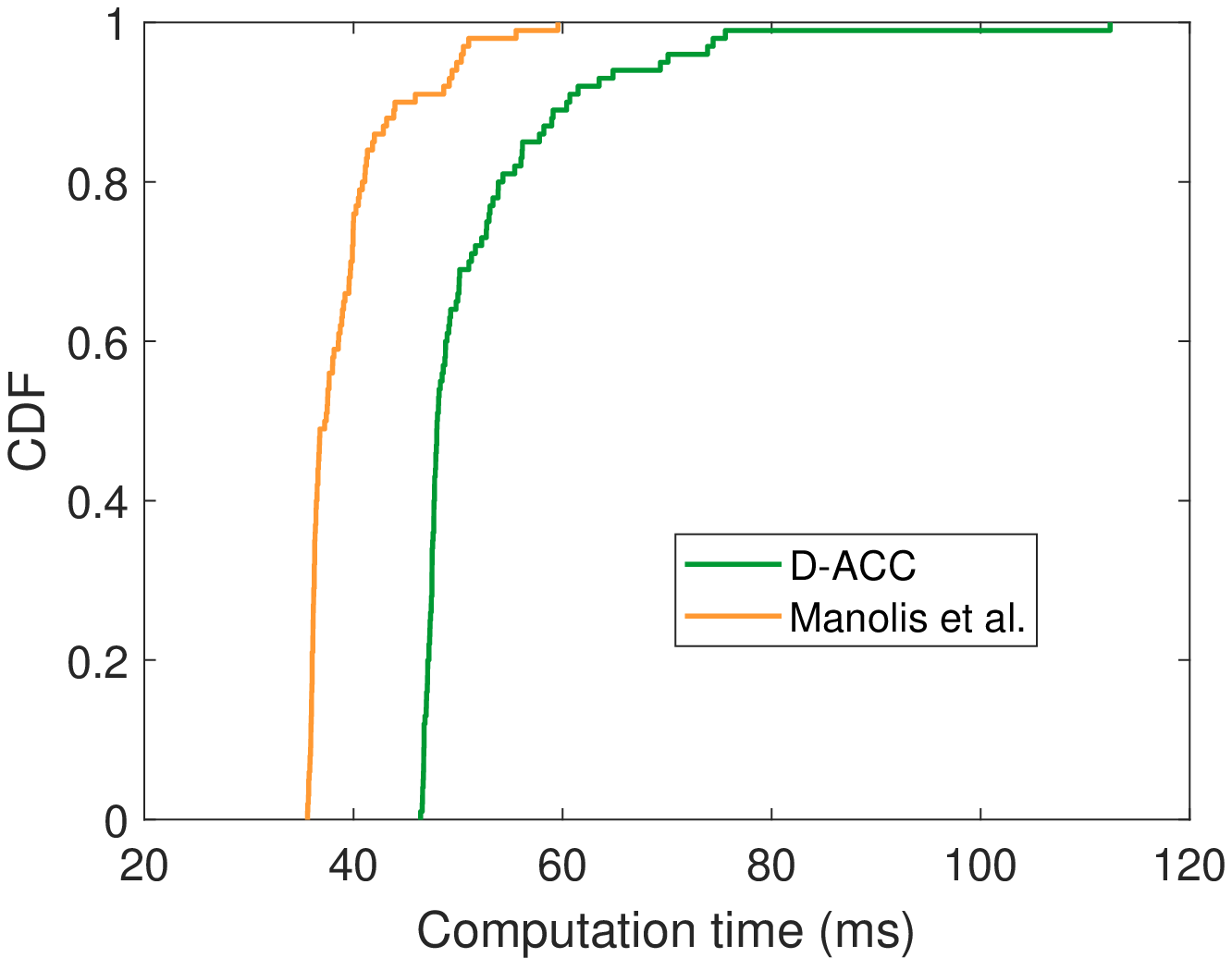}
    \caption{Computation time. \label{fig:computation_time}}
\end{minipage}
\end{figure}

We evaluate the effect of the traffic density of the main lane on the performance of D-ACC. We measure the average vehicle speed by varying the traffic density. Fig.~\ref{fig:main_lane_density} shows the results. As expected, as the vehicle density increases, the average speed for both D-ACC and Manolis \emph{et al.} decreases. We observe that D-ACC achieves higher average speed than Manolis \emph{et al.} by up to 14.6\% especially when the vehicle density is low. An interesting observation is that as the main lane traffic is getting more saturated, the performance gain of D-ACC compared to Manolis \emph{et al.} becomes smaller because merging vehicles cannot change lanes easily when the traffic density is too high.


\subsection{Computation Time}
\label{subsec:computation}


We measure the computation time for obtaining the headway distance for both D-ACC and Manolis \emph{et al.} to demonstrate that D-ACC is capable of adjusting the headway distance quickly in response to dynamically changing traffic conditions. Specifically, the computation time for each vehicle is measured 100 times for both D-ACC and Manolis \emph{et al.} Fig.~\ref{fig:computation_time} depicts the cumulative distribution function (CDF) graph for the results. Here we assume that the neural network model for D-ACC has been already established. The average computation time for D-ACC is 64.5\% higher than that for Manolis \emph{et al.} partly due to the computation overhead for taking into account various factors for determining the headway distance. However, we note that the average computation time of 51.7ms for D-ACC is sufficient to make a decision at every update interval which is one second.

\section{Conclusion}
\label{sec:conclusion}

We have presented a dynamic adaptive cruise control system (D-ACC) designed to adjust the headway distance adaptively based on reinforcement learning in response to dynamically changing traffic conditions for highways with ramps. We demonstrated that D-ACC improves traffic flow significantly even with a small penetration rate compared with a state-of-the-art real-time ACC system. Our on-going work is to develop a smartphone-based D-ACC, integrate it with an actual vehicle platform and evaluate the performance in real-world traffic environments by recruiting vehicles through crowdsourcing websites such as Amazon Mechanical Turk.

\bibliographystyle{IEEETran}
\bibliography{mybibfile}

\end{document}